\documentstyle[epsfig]{aipproc}

\def\aj{Ast. J.}
\def\apj{Ap. J.}
\def\apjl{Ap. J. (Letters)}

\def\araa{Ann. Rev. Astr. Ap.}
\def\nat{ Nature}

\def\mnras{ MNRAS}

\def\stis{{\em STIS}}
\def\hst{{\em HST}}

\def\hnot{H$_0$}
\def\qnot{q$_0$}
\def\msun{\ifmmode {\rm M_\odot} \else M$_\odot$\fi}
\def\lsun{\ifmmode {\rm L_\odot} \else L$_\odot$\fi}

\def\kms{km s$^{-1}$}
\def\deg{\ifmmode ^{\circ}
         \else $^{\circ}$\fi}
\def\pdeg{\ifmmode 
           $\setbox0=\hbox{$^{\circ}$}\rlap{\hskip.11\wd0 .}$^{\circ}
     \else \setbox0=\hbox{$^{\circ}$}\rlap{\hskip.11\wd0 .}$^{\circ}$\fi}

\def\arcsec{\ifmmode '' \else $''$\fi}
\def\arcsecpt{\ifmmode ''\!\!. \else $''\!\!.$\fi}

\def\msunyr{\ifmmode {\rm M_\odot~yr^{-1}}\else${\rm M_\odot~yr^{-1}}$\fi}
\def\lam{\ifmmode {\lambda} \else {$\lambda$} \fi}

\def\mdoto{\ifmmode {\dot{M}_0} \else  $\dot{M}_0$ \fi}
\def\teff{\ifmmode {T_{eff}} \else $T_{eff}$ \fi}
\def\ilam{\ifmmode {I_\lambda} \else  $I_\lambda$ \fi}
\def\inu{\ifmmode {I_\nu} \else  $I_\nu$ \fi}
\def\fnu{\ifmmode {F_\nu} \else  $F_\nu$ \fi}

\def\yr{\ifmmode {\rm yr} \else  yr \fi}
\def\cm{\ifmmode {\rm cm} \else  cm \fi}
\def\cmmitwo{\ifmmode \rm cm^{-2} \else $\rm cm^{-2}$\fi}
\def\cmmithree{\ifmmode \rm cm^{-3} \else $\rm cm^{-3}$\fi}
\def\cmps{\ifmmode \rm cm~s^{-1}\else $\rm cm~s^{-1}$\fi}
\def\cmpsps{\ifmmode \rm cm~s^{-2}\else $\rm cm~s^{-2}$\fi}
\def\kmps{\ifmmode \rm km~s^{-1}\else $\rm km~s^{-1}$\fi}
\def\kmpspmpc{\ifmmode \rm km~s^{-1}~Mpc^{-1} \else
    $\rm km~s^{-1}~Mpc^{-1}$\fi}
\def\ergps{\ifmmode \rm erg~s^{-1} \else $\rm erg~s^{-1}$ \fi}
\def\ergpspcm{\ifmmode \rm erg~s^{-1}~cm^{-2} \else $\rm erg~s^{-1}~cm^{-2}$ \fi}
\def\ergpspcmphz{\ifmmode \rm erg~s^{-1}~cm^{-2}~Hz^{-1} \else $\rm
erg~s^{-1}~cm^{-2}~Hz^{-1}$ \fi}
\def\ergpspcmpa{\ifmmode \rm erg~s^{-1}~cm^{-2}~\AA^{-1} \else $\rm
erg~s^{-1}~cm^{-2}~\AA^{-1}$ \fi}
\def\ergpsphz{\ifmmode \rm erg s^{-1} Hz^{-1} \else 
   $\rm erg s^{-1} Hz^{-1}$ \fi}

\begin{document}

\title{Discovery of a 2 Kpc Binary Quasar
\thanks{Based on observations made with the NASA/ESA Hubble
Space Telescope.  STScI is operated by the Association of Universities
for Research in Astronomy, Inc. under NASA contract NAS5-26555.}
}

\author{G. A. Shields$^*$,
V. Junkkarinen$^{\dagger}$
\thanks{Visiting Astronomer, Cerro Tololo Inter--American
Observatory, which is operated by the Association of Universities
for Research in Astronomy, Inc., under contract with the National
Science Foundation.},
E. A. Beaver$^{\dagger}$,
E. M. Burbidge$^{\dagger}$,
R. D. Cohen$^{\dagger}$,
F. Hamann$^{\ddagger}$, and
R. W. Lyons$^{\dagger}$}
\address{$^*$Department of Astronomy, University of Texas, Austin TX
78712\\
$^{\dagger}$Center for Astrophysics and Space Sciences,
University of California, San Diego\\ La Jolla, CA 92093\\
$^{\ddagger}$Department of Astronomy, University of Florida, 
Gainsville FL 32611}

\maketitle

\begin{abstract}

LBQS 0103$-$2753  
is a binary quasar with a separation of only 0.3 arcsec.  The projected
spacing of 2.3 kpc at the distance of the source (z = 0.848) is much smaller
than that of any other known binary QSO. The binary nature is demonstrated by
the very different spectra of the two components and the low probability of a
chance pairing.  LBQS 0103$-$2753 presumably is a galaxy merger with
a small physical separation between the two supermassive black holes. Such
objects may provide important constraints on the evolution of binary black
holes and the fueling of AGN.

\end{abstract}

\section*{Introduction}

Binary QSOs occur at a rate of about one per thousand QSOs
\cite{hewett98,kochanek99}.  Typical angular separations are 3\arcsec\ to
10\arcsec.  True binaries, as opposed to gravitational lenses, may be
recognized by different radio emission or different emission-line or BAL
characteristics in their component spectra 
\cite{kochanek99,mortlock99}.  Binary QSOs presumably
represent galaxy encounters in which tidal interactions have led to an enhanced
probability of nuclear activity \cite{djorgovski91}.  Such events may offer
important insights into the fueling of AGN and the timescales for the mergers
of galaxies and their central black holes \cite{begelman80}.

In the course of a study of QSOs with broad absorption lines (BALs), we have
discovered that LBQS 0103$-$2753 is a binary QSO with a separation of only
0\arcsecpt3  \cite{junkkarinen01}.  This corresponds to a physical
separation of only 2 kpc.  The closest previously known binaries have spacings
of 15 kpc or more  \cite{brotherton99}.  Such a close spacing suggests
an advanced merger, with the two supermassive black holes well within
the nascent bulge of the merged galaxy.

\section*{Observations}

LBQS 0103$-$2753  is one of eight BALQSOs included in a study,
by the authors, of the ultraviolet spectra of BALQSOs at moderate redshift with
the Space Telescope Imaging Spectrograph (\stis) on the Hubble Space Telescope
(\hst).   The object has an emission-line redshift $z_e$ = 0.848 and an
apparent magnitude $B_J$ = 18.1, and its LBQS spectrum shows broad absorption
in Mg II $\lambda$2798  \cite{morris91}.  Figure 1
shows the STIS CCD acquisition image.  The A/B flux ratio is 3.1, implying
magnitudes $B_J$ = 18.2 and 19.4 for components A and B,
respectively.  The FWHMs of the images are consistent with point sources, and
the separation is 
$\Delta\theta$=0.295 $\pm$ 0.011 arcsec in position angle +30.1$\deg$ $\pm$
2.2$\deg$ (A northeast of B).


\begin{figure}[b!] 
\centerline{\epsfig{file=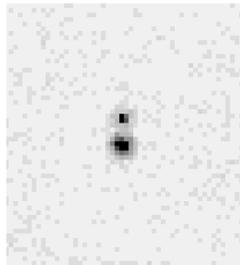,height=2.0in}}
\vspace{10pt}
\caption{Acquisition image of LBQS 0103$-$2753 obtained with
the STIS CCD.
The brighter object is A, the BALQSO, and the fainter is B.
Reproduced from Junkkarinen et al. (2001) with permission.}
\label{fig1}
\end{figure}

A 52\arcsec\ by 0\arcsecpt2\ slit was used for the spectroscopic observations
(Figure 2) with the STIS NUV--MAMA using the G230L grating.
The QSO pair was, by luck, almost perfectly aligned with the STIS slit,
itself aligned along the image +y axis.  
The resolution at 2400~\AA\ is 3.3~\AA\ FWHM.

\begin{figure}[b!] 
\vspace{-15pt}
\centerline{\epsfig{file=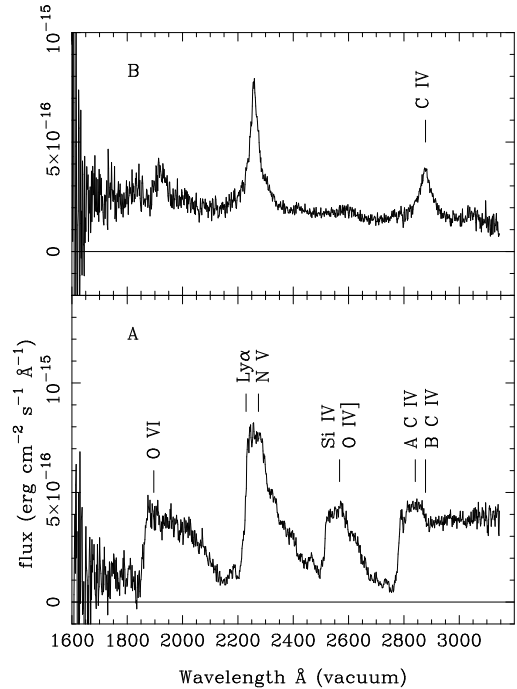,height=4in}}
\caption{Spectra of LBQS 0103$-$2753 A and B
obtained with HST/STIS using the NUV--MAMA and G230L grating.
The upper panel is
component B and the lower panel is component A.
The measured C IV $\lambda$ 1549 emission-line wavelengths for components A and
B are both shown with vertical marks above component A's CIV
$\lambda$ 1549 emission line.
Reproduced from Junkkarinen et al. (2001) with permission.}
\label{fig2}
\end{figure}

We obtained an optical spectrum of LBQS 0103$-$2753 (sum of both components)
in August 1998 using the CTIO Blanco 4m telescope.
The spectrum  \cite{junkkarinen01} shows a dramatic weakening of the
Mg II Bal, which is unusual
 \cite{junkkarinen99}.

\section*{LBQS 0103$-$2753: a Close Binary QSO}

LBQS 0103$-$2753 A has strong BALs of 
C IV $\lambda\lambda$1548,1551, Si IV $\lambda$1400,
N V $\lambda$1240, and O VI $\lambda$1034 that are absent in the spectrum of
component B. In addition, the two components have very different emission-line
properties.  This effectively rules out the possibility that the object is a
lensed QSO.  Junkkarinen et al. \cite{junkkarinen01} argue that the likelihood
also is small that LBQS 0103$-$2753 is a chance alignment of a wider binary of
the 3\arcsec\ to 10\arcsec\ ($\sim$40~kpc) variety.

Could LBQS 0103$-$2753 be a chance superposition of two unrelated QSOs?  The
emission-line redshifts of the two components, though similar, differ by more
than the 600~\kms\ difference typical of binary QSOs  \cite{kochanek99}. 
The redshift of B is well determined at $z_B$ = 0.858, but the BALs
and emission-line profiles of
component A make the emission-line redshift quite uncertain.  The C IV line
gives $z_A$ = 0.834, which corresponds to a velocity difference from $z_B$ of
3900~\kms. However, C~IV emission lines are often
blueshifted from the systemic redshift
 \cite{espey89}, especially in the case of weak, low contrast
lines as in component A  \cite{brotherton94}. 
How likely is a chance coincidence of a QSO as bright as the apparent magnitude
of component B within this redshift range and within an
angular separation of 0\arcsecpt3?   We take a
surface density of 10 QSOs per square degree between B = 18.2 and 19.4, for
QSOs with redshifts $z < 2.2$ \cite{boyle88}.  If we assume these are
distributed roughly uniformly with redshift, then the odds of having a second
QSO within 0\arcsecpt3 of a given QSO, and with a redshift difference not more
than
$\Delta z = z_B - z_A = 0.024$, is
$10^{-8.6}$.  If $\sim$500 QSOs have been observed in a way that would reveal
such a binary (see below), then the odds of having one chance
alignment resembling LBQS 0103$-$2753, among all these targets, is only
$10^{-5.9}$.  This supports the conclusion that LBQS 0103$-$2753 is a true,
physical binary.

\section*{Discussion}

The greatest number of opportunities to discover a pair like LBQS 0103$-$2753
appears to be offered by the
\hst\ snapshot survey for gravitationally lensed QSOs \cite{maoz93}. 
This involved 498 QSOs with
$z > 1$ and $M_V < -25.5$ (\hnot=100, \qnot = 0.5). 
The snapshot survey could have detected a binary with the
characteristics of LBQS 0103$-$2753 \cite{bahcall92}.  However, it
did not find any binary QSOs with separations less than 1\arcsecpt0.
From the one example of LBQS 0103$-$2753, we therefore estimate a rate of
roughly 1/500 for separations $\sim$0\arcsecpt3. This in turn suggests that
there are $\sim$10 unrecognized binaries in the 0\arcsecpt3 range among the
$\sim$$10^4$ known QSOs.

Junkkarinen et al. \cite{junkkarinen01} suggest that LBQS 0103$-$2753
represents an advanced stage of a galactic merger.  The cores of the two
galaxies are merging, and the black holes are settling into an orbit of radius
similar to the observed spacing of 2 kpc.  The wider binaries observed, with
physical spacing of $\sim$20 to 100 kpc, likely represent mergers at an earlier
stage.  Simulations of colliding disk galaxies \cite{barnes96}
indicate that gas 
accumulates in the nucleus of each galaxy as the two orbit away from each
other after the first close encounter.  The duration of this first large orbital
loop is
$\sim$$10^{8.5}$ yr. Dynamical friction causes rapid decay of subsequent
loops, and the central regions of the
galaxies merge quickly. The final merger involves a massive concentration of gas
in the nucleus capable of fueling powerful starbursts and AGN activity. 
Junkkarinen et al. \cite{junkkarinen01} estimate the masses of the black holes
and the properties of the host galaxies from the QSO's luminosities and the 
relationship between black hole mass and bulge properties of the host galaxy
\cite{gebhardt00}.  This leads to an estimated lifetime of the binary
black hole orbit, at the observed 2 kpc spacing, of only $\sim$$10^7$ years.
This in turn suggests that the probability of a merger being active as a binary
QSO is larger during the 2 kpc stage than during the 40 kpc stage.

Ultraluminous infrared galaxies (ULIGS, $L_{ir} >
10^{11}~\lsun$) typically involve galactic mergers  \cite{sanders96,genzel00}. 
Interestingly, the mean nuclear spacing of
$\sim$2 kpc is similar to that of LBQS 0103$-$2753.  
The brighter ULIGs are believed to be powered predominantly by AGN.  LBQS
0103$-$2753 is distinguished from typical ULIGs by the lack of heavy
extinction, as well as an exceptionally high luminosity.  Infrared observations
of LBQS 0103$-$2753 would be valuable to determine whether massive amounts of
dust are still present in the vicinity of the QSOs.

\section*{Other Unresolved Binary QSOs?}

We have estimated that binaries as close as LBQS 0103$-$2753 occur with a
frequency of
$\sim$$10^{-3}$.  Given this low incidence, searches for 0\arcsecpt3 binaries
would benefit from any criterion to narrow the list of candidates.  Until our
\hst\ observations, LBQS 0103$-$2753 was not, to our knowledge, suspected of
being a binary QSO.  The great diversity of QSO spectra makes it difficult to
specify spectroscopic criteria with which to identify candidate binaries. 
Objects with double peaked narrow line emission ([O III]~$\lambda$5007 
or [O II]~$\lambda$3727) might include some candidates.  Composite broad
emission-line profiles of an unusual character might also be an indicator, e.g.,
the QSO B340  \cite{shields81}.  Ideally, one could identify
two properties of QSOs that almost never occur together. The rare exceptions
would then be candidates for unresolved binaries.  One possibility might be
BALQSOs with strong, unabsorbed soft X-ray fluxes; BALQSOs as a class show weak
soft X-ray fluxes, relative to the optical continuum, evidently because of
absorption  \cite{brandt00}.

Support for this work was provided by NASA through grants
GO-07359, GO-07359.02 from the Space Telescope Science Institute,
which is operated by the Association of Universities for Research in
Astronomy under NASA contract NAS5-26555.

\end{document}